# Non Uniform On Chip Power Delivery Network Synthesis Methodology

Patrick Benediktsson
Institution of Telecomunications,
University of Lisbon, Portugal

Jon A. Flandrin
Institution of Telecomunications,
University of Lisbon, Portugal

Chen Zheng
Intel Corp.
USA

**Abstract**
**In this paper, we proposed a non-uniform power delivery network (PDN) synthesis methodology. It first constructs initial PDN using uniform approach. Then preliminary power integrity analysis is performed to derive IR-safe candidate window. Congestion map is obtained based global route congestion estimation. A self-adaptive non-uniform PDN synthesis is then performed to globally and locally optimize PDN over selected regions. The PDN synthesis is congestion-driven and IR-guarded. Experimental results show significant timing important in trade-off small PDN length reduction with no EM/IR impact. We further explored potential power savings using our non-uniform PDN synthesis methodology.**

**Keywords — power delivery network, non-uniform, power integrity, congestion**

## 1. Introduction

Morden semiconductor integrated circuit design has been scaling drastically into nanometer feature size range. To achieve better performance, higher power density and more routing resource is required. Both post challenges to on-chip power delivery network (PDN) design, which needs to deliver power supply to every transistor on die. The PDN needs to meet electromigration (EM) and IR limit to guarantee power integrity of a chip [1]. EM limit requires current density through each PDN wire segment and via to be lower than a threshold value, while IR limit requires the voltage drop when reaching transistor source node. Failure to satisfy power integrity will lead to reliability issues or chip malfunction [2]. Meanwhile, wire RC delay has become increasingly significant and dominants the propagation delay of a large portion of critical paths [3]; therefore, either more and more routing layers are added to increase routing resource or PDN resource needs to be carefully assigned to save space for signal routing. Thus, efficient PDN design and optimization is the key to deliver successful higher performance chip in current and future technology nodes.

In this paper we propose a non-uniform PDN design methodology that optimizes PDN both globally and locally that occupies less routing resource compared to traditional PDN design.
The saved routing space helps to improve signal routing in congested area and achieve better timing results. Further exploration shows that potential power saving can be obtained by leakage reduction through cell threshold voltage optimization.

## 2. Motivation

In traditional PDN design, designers first estimate the chip power density, current density and IR drop based on technology information, such as cell power, sheet resistance, etc. Then a uniform PDN is designed to meet the power integrity requirement based on estimation. A few iterations are performed to modify the initial version to eventually close EM/IR limit [4]. PDN design and optimization technique has been widely investigated by researchers. In [5], S. Kose proposed a power grid design based on effective resistance. Z. S. Zheng provided a tradeoff optimization with voltage regulation in [6]. While in [7], T. Hayashi investigated power grid optimization algorithm based on manufacturing cost restriction. However, there are a few limitations to previous conventional approach: (1) since cell density is not uniformly distributed, the power density estimation is usually over pessimistic to cover possible local power hotspots; (2) since PDN is designed first before place and route happens, PDN designers have no knowledge how the actual power distribution look like, thus it is impossible to apply non-uniform style PDN to optimize for different design blocks. Figure 1 shows a typical IR drop distribution and how PDN designed for high percentile power density could be over-designed for low percentile power density. Often, temperature and variation effects are included in the analysis to obtain more realistic results [9]. The two limitations implies inevitable over-design of traditional style PDN. To overcome this shortage, we propose an EM/IR-aware and congestion-driven non-uniform PDN design methodology that globally meets power integrity requirement and locally optimizes signal routing resource. To our best knowledge, this is the first systematic work to address the non-uniform style PDN synthesis.

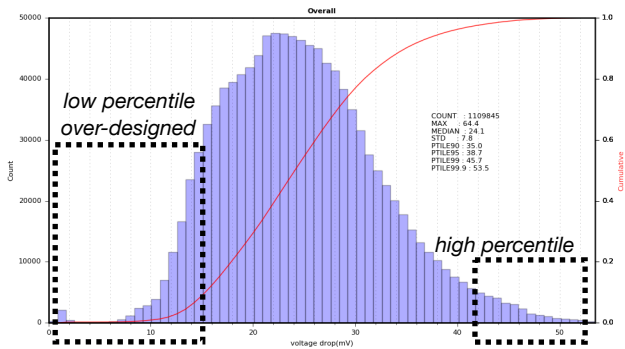

**Fig 1. Typical IR distribution**

The non-uniform PDN methodology starts with traditional uniform PDN construction. The uniform PDN is fed into regular place, clock synthesis and post-clock optimization. Then a preliminary EM/IR analysis is run on a pre-route database. Based on the analysis results, we identify design regions that are IR safe as potential non-uniform PDN modification candidate. A routing congestion map based on global route engine is used to drive the actual non-uniform PDN synthesis. In later sections, we discuss in detail the candidate selection, non-uniform PDN synthesis and timing/power benefits using our proposed methodology.

## 3. Candidate Selection

In our approach, the entire design is divided into unit size windows (e.g. 20um x 20um). Preliminary EM/IR result and routing congestion map for each window is derived and used as basic metrics for candidate selection.

The reasoning for running a preliminary EM/IR analysis on a pre-route database is that cell placement are largely determined by the time post-clock optimization is done, so the result correlates well with the final sign-off EM/IR analysis. It is critical to avoid reduce PDN over EM/IR critical regions as reliability is a key factor to guarantee quality design [8].Besides, although non-uniform PDN synthesis can be introduced after route, during engineering-change-order (ECO) stage, it will be best to interfere before route stage for router to best leverage its benefits. Later in this section, we will discuss how fast IR analysis and congestion estimation is done for each unit size window.

### 3.1 Fast IR analysis, guard-band window

To determining if a window is IR-safe, we use single hotspot filtering and average margin threshold mechanism.

Due to the continuity of voltage drop distribution [10], it is not sufficient to claim a window is IR-safe even if all instances voltage drop is below required limit. For example, if window A is IR-safe, while window B adjacent to window A is potentially IR-sensitive, then modifying PDN over window A may affect window B and cause IR failure in it. To address this concern, we introduce guard-band window, which is a larger window that extends beyond the original unit size window. In our experiment, we found a guard-band length that is same as unit step serves the purpose of estimating IR-drop margin well. Figure 2 illustrate how guard-band window and unit size window work together to identify IR-safe candidate.

### 3.2 Congestion Estimation

Since our target is to reduce PDN occupancy and give back resource to signal routing, it is important to estimate the routing congestion within a given unit size window.

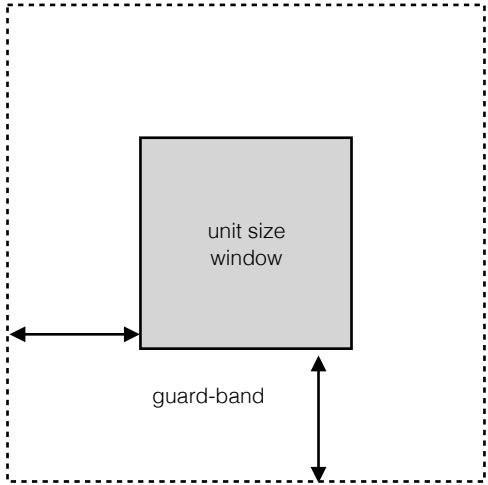

**Fig 2. Guard-band window**

Global router will start with a minimum spanning tree (MST) and converted it to a steiner tree [11]. Depending on the horizontal/vertical traffic and capacity across all layers, routing congestion is estimated. Figure 3 illustrates the global route congestion estimation model. High pin density usually causes routing congestion hotspot and result in overflow issue. Detailed router either needs to detour or to end up with high density routing that introduces more parasitic loads and causes cross-talk noise issue. Reducing PDN occupancy over those regions will help improve the routing results.

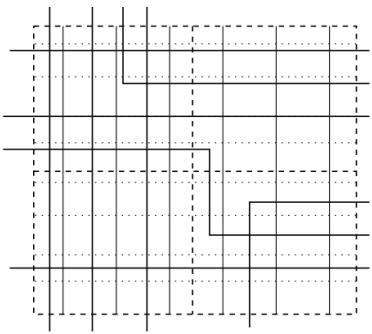

**Fig 3. Global route congestion**

In our experiments, we rely on the global router congestion map to select the candidate for non-uniform PDN synthesis for most benefits gain.

## 4. Non-Uniform PDN Synthesis

The non-uniform PDN synthesis is a self-adaptive process. It works on the selected candidate windows, and calculates the ratio of PDN reduction that needs to be applied to the given candidate window. We define a target function F for the non-uniform PDN synthesis:

$$F = \alpha \cdot \text{congestion} + \beta \cdot \frac{1}{\text{IR drop}} \quad (0 \leq F \leq 50\%) \quad (1)$$

To avoid over optimistic PDN reduction, we set the maximum allowed reduction rate to be 50%. It is necessary to leave certain margin as variation effect can degrade EM/IR results significantly [12]. In equation (1), $\alpha$ is the coefficient for congestion score and $\beta$ is the weight for IR-drop. Equation (1) implies the congestion-driven and IR-drop guarded characteristic of the proposed non-uniform PDN synthesis. Figure 4 illustrates how the non-uniform PDN synthesis works on a candidate window. Due to the redundancy property of PDN, we are able to maintain the power integrity with non-uniform PDN synthesis [13]. It should be noted that the PDN reduction should not impact regular signal wires as their EM-sensitivity is low [14]. In [15], a systematic solution to achieve EM reliable design is also provided.

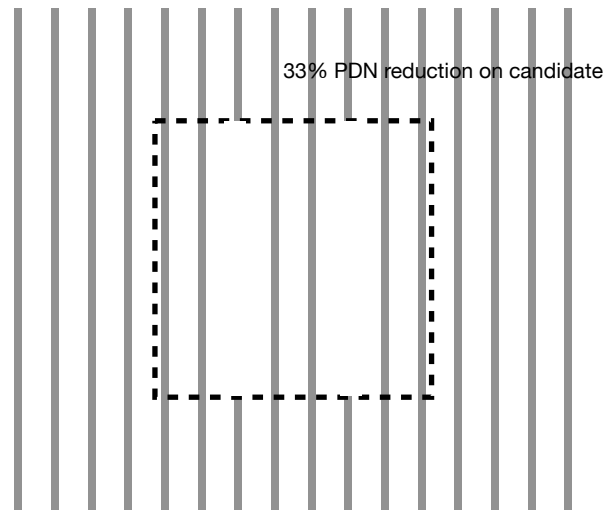

**Fig 4. Non-uniform PDN synthesis**

## 5. Experimental Results

We used two blocks from the open source design or1200_fpu_fcmp and or1200_genpc [16]. Physical design is implemented using Synopsis ICC2 [17]. We used RedHawk [18] for preliminary EM/IR analysis. True worst case current analysis [19] is also included to account for EM analysis. After non-uniform PDN synthesis, layout aware analysis [20] is performed to ensure power integrity is not affected. Figure 5 shows the IR distribution before and after the non-uniform PDN synthesis. It shows that IR impact from non-uniform PDN synthesis is minimal and the outlier part remains the same, thus no extra effort is introduced on designer's side to close IR. It also implies EM reliability is satisfied as voltage drop and current density is closely correlated [21]. Table 1 shows the PDN length reduction on some of the metal layers. The larger amount of PDN length reduction on vertical layers indicate that the design the vertically congested. Table 2 shows that timing improvement after route between baseline and the proposed non-uniform PDN synthesis. Notable improvements are marked with gray. It can be observed that with small portion of PDN reduction, timing analysis results can improve significantly. This is because the congestion-driven non-uniform PDN synthesis targets at the critical windows which give the most benefits gain.

Besides timing benefits, the non-uniform PDN synthesis can also provide potentially power benefits. One such example is to remove the congestion-driven feature, but brute-force apply the non-uniform PDN synthesis. By doing this, some positive paths can have chance to swap certain cells to higher threshold voltage, thus saving leakage power. Table 3 shows cell count for different threshold voltages and overall power savings.

## 6. Conclusions

In this paper, we discuss the limitations of traditional uniform PDN synthesis. The uniform PDN is usually pessimistic and over-designed. We propose a self-adjust non-uniform PDN synthesis methodology. It uses preliminary IR analysis and congestion map as candidate selection criteria.

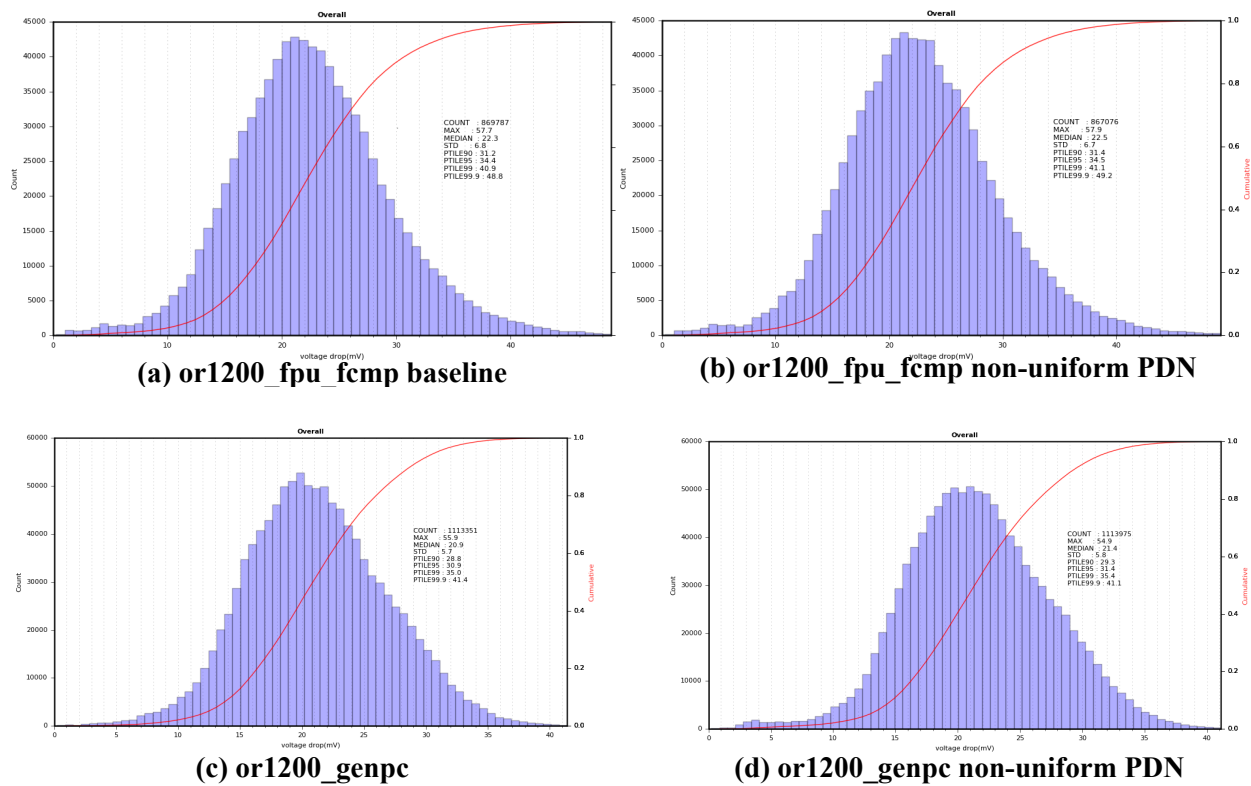

**Fig 5. Comparison of non-uniform PDN synthesis**

Table 1. PDN length comparisons

| PDN length | or1200_fpu_fcmp | | | or1200_genpc | | |
|---|---|---|---|---|---|---|
| | base | non-uniform | delta | base | non-uniform | delta |
| M2 | 7357.763 | 7317.763 | 0.00% | 649286.5775 | 597906.5775 | 7.91% |
| M3 | 336815.783 | 315505.463 | 6.33% | 264770.718 | 238230.718 | 10.02% |
| M4 | 193545.941 | 190465.941 | 1.42% | 159921.1195 | 157821.1195 | 1.31% |
| M5 | 139799.8615 | 130459.8615 | 5.99% | 114523.972 | 105023.972 | 8.30% |
| M6 | 214127.1025 | 211027.1025 | 1.43% | 160059.527 | 157959.527 | 1.31% |
| M7 | 167574.516 | 158114.516 | 6.03% | 117514.9265 | 107894.9265 | 8.19% |

Guard-band window is introduced to minimize IR impact. Target function is defined to self-adaptively reduce PDN over candidate windows. Results show promising timing improvement with zero IR impact. We further explored potential power saving with brute-force PDN reduction. Results show average of 2.95% total power saving. As power deliver network has become more and more challenge to design and close power integrity as well as save sufficient space for routing, we believe there are more optimization space for non-uniform PDN design that can be explored further. This technique can also combine with other existing power grid optimization

Table 2. Comparison summary

| | fpu base | fpu exp | delta | pc base | pc exp | delta |
|---|---|---|---|---|---|---|
| TNS | 93.147 | 78.916 | 15.28% | 423.535 | 384.095 | 9.31% |
| TNS (REG) | 37.675 | 26.202 | 30.45% | 274.671 | 247.898 | 9.75% |
| MAX (IR) | 64.4 | 64.3 | | 57.7 | 57.9 | |
| MEDIAN | 24.1 | 24.8 | | 22.3 | 22.5 | |
| PTILE90 | 35.0 | 35.5 | | 31.2 | 31.4 | |

Table 3. Power comparisons

| | fpu base | fpu exp | pc base | pc exp |
|---|---|---|---|---|
| SVT | 498872 | 598298 | 240943 | 282149 |
| LVT | 309843 | 323078 | 118274 | 135539 |
| ULVT | 573954 | 535897 | 259384 | 224948 |
| ELVT | 15493 | 13049 | 10394 | 6525 |
| Power | 1784.3 | 1712.9 | 432.7 | 408.9 |

techniques [22-25] to further improve PDN design.